# A comprehensive survey on computational learning methods for analysis of gene expression data

**Nikita Bhandari[1], Rahee Walambe[2,3], Ketan Kotecha[1,3], Satyajeet P. Khare[4]**

[1] Computer Science Department, Symbiosis Institute of Technology, Symbiosis International (Deemed University), Pune, MH, India

[2] Electronics and Telecommunication Department, Symbiosis Institute of Technology, Pune, Maharashtra, India

[3] Symbiosis Center for Applied AI (SCAAI), Symbiosis International Deemed University, Pune, Maharashtra, India

[4] Symbiosis School of Biological Sciences, Symbiosis International Deemed University, Pune, Maharashtra, India

**\* Correspondence:**

Rahee Walambe: rahee.walambe@sitpune.edu.in

Ketan Kotecha: drketankotecha@gmail.com

Satyajeet Khare: satyajeet.khare@ssbs.edu.in

**Abstract**

Computational analysis methods including machine learning have a significant impact in the fields of genomics and medicine. High-throughput gene expression analysis methods such as microarray technology and RNA sequencing produce enormous amounts of data. Traditionally, statistical methods are used for comparative analysis of gene expression data. However, more complex analysis for classification of sample observations, or discovery of feature genes requires sophisticated computational approaches. In this review, we compile various statistical and computational tools used in analysis of expression microarray data. Even though the methods are discussed in the context of expression microarrays, they can also be applied for the analysis of RNA sequencing and quantitative proteomics datasets. We discuss the types of missing values, and the methods and approaches usually employed in their imputation. We also discuss methods of data normalization, feature selection, and feature extraction. Lastly, methods of classification and class discovery along with their evaluation parameters are described in detail. We believe that this detailed review will help the users to select appropriate methods for preprocessing and analysis of their data based on the expected outcome.

## 1 Introduction

A genome is a complete set of genes in an organism. Genomics is a study of the information structure and function programmed in the genome. Genomics has applications in multiple fields, including medicine (Chen *et al.*, 2018; Lai *et al.*, 2020; Huang *et al.*, 2021), agriculture (Abberton *et al.*, 2016; Parihar *et al.*, 2022), industrial biotechnology (Alloul *et al.*, 2022), synthetic biology (Baltes and Voytas, 2015), etc. Researchers working in these domains create and use a variety of data such as DNA, RNA, and protein sequences, gene expression, gene ontology, protein-protein interactions (PPI), etc.

Genomics data can be broadly classified into sequence and numeric data (e.g., gene expression matrix). The DNA sequence information can be determined by first generation (Sanger, Nicklen and Coulson, 1977), second generation sequencing (Margulies *et al.*, 2005; Shendure *et al.*, 2005; Bentley *et al.*, 2008; Valouev *et al.*, 2008) or third generation sequencing (Harris *et al.*, 2008; Eid *et al.*, 2009; Eisenstein, 2012; Rhoads and Au, 2015) methods. The second and third generation sequencing are together referred to as Next Generation Sequencing (NGS). Applications of DNA sequence analysis include prediction of protein sequence and structure, molecular phylogeny, identification of intrinsic features, sequence variations, etc. Common implementations of these applications include splice site detection (Nguyen *et al.*, 2016; Fernandez-Castillo *et al.*, 2022), promoter prediction (Umarov and Solovyev, 2017; Bhandari *et al.*, 2021), classification of diseased related genes (Peng, Guan and Shang, 2019; Park, Ha and Park, 2020), identification of protein binding sites (Pan and Yan, 2017; Uhl *et al.*, 2021), biomarker discovery (Arbitrio *et al.*, 2021; Frommlet *et al.*, 2022), etc. The numeric data often generated from functional genomics studies include gene expression, single nucleotide polymorphism (SNP), DNA methylation, etc. Microarray and NGS technologies are the tools of choice for functional genomics studies. The functional genomics that deals with high-throughput study of gene expression is referred to as transcriptomics.

Gene expression data, irrespective of the platform used (e.g., microarray, NGS, etc.), contains the expression levels of thousands of genes experimentally evaluated in various conditions. Gene expression analysis helps us understand gene networks and molecular pathways. Gene expression information can be utilized for basic as well as clinical research (Behzadi, Behzadi and Ranjbar, 2014; Chen *et al.*, 2016; Karthik and Sudha, 2018; Kia *et al.*, 2021). In disease biology, gene expression analysis provides an excellent tool to study the molecular basis of disease as well as the identification of markers for diagnosis, prognosis, and drug discovery. Therefore, for this review, we will focus on computational methods in the analysis of gene expression data.

The data produced by microarray as well as NGS-based RNA sequencing goes through multiple phases of quality check before analysis. This data is further transmuted to a numerical matrix (Figure 1) where rows and columns represent genes and samples. The numeric value in each cell of a matrix links the expression level of a specific feature gene to a particular sample. The expression matrix is generally a flat dataset as the number of features is very high compared to the number of samples. Some of the standard DNA microarray platforms available are Affymetrix (Pease *et al.*, 1994), Agilent (Blanchard, Kaiser and Hood, 1996), etc. Some of the standard commercial NGS platforms are Illumina (Bentley *et al.*, 2008), Ion torrent (Rothberg *et al.*, 2011) etc. The massive amount of data generated from publicly funded research is available through open access repositories such as Gene Expression Omnibus (GEO), ArrayExpress, Genomic Expression Archive (GEA), etc. (Table 1).

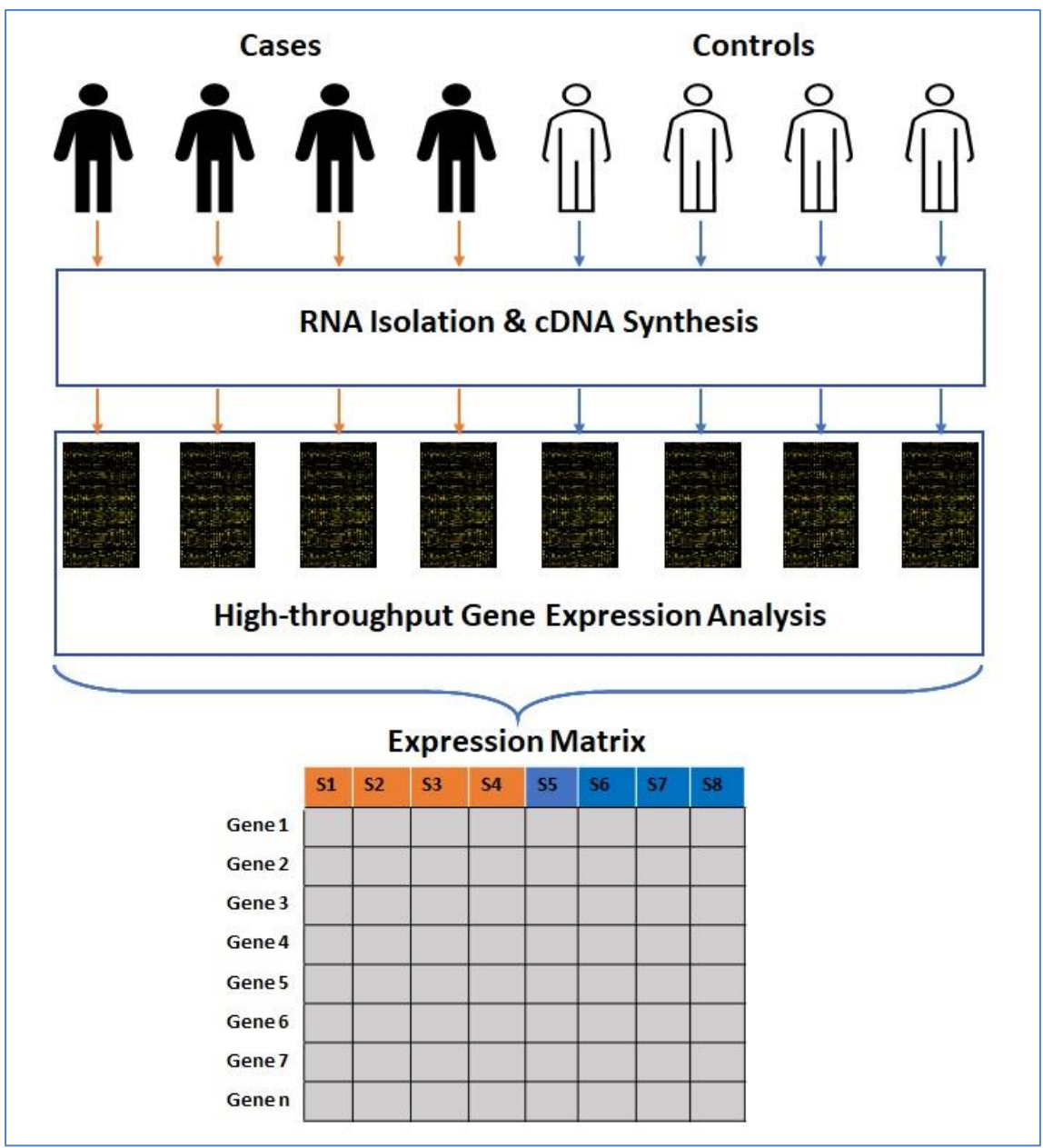

**Figure 1: Process of generation of high-throughput gene expression data.** The clinical samples are subjected to RNA isolation and cDNA synthesis. The cDNAs are subjected to high-throughput gene expression analysis. The raw data obtained from these methods is further transmuted into a numerical matrix where rows and columns represent genes and samples.

Identification of differentially expressed genes is the most common application in gene expression analysis. This type of class comparison analysis can be achieved using basic statistical techniques, for example, chi-squared test, t-test, ANOVA, etc. (Segundo-Val and Sanz-Lozano 2016). Commonly used packages for microarray-based gene expression analysis include limma (Smyth, 2005), affy (Gautier *et al.*, 2004), lumi (Du, Kibbe and Lin, 2008), oligo (Carvalho and Irizarry, 2010); whereas, those for RNA sequencing analysis include EdgeR (Robinson, McCarthy and Smyth, 2009) and DESeq2 (Love, Huber and Anders, 2014). The classification and regression problems on the other hand depend on classical linear and logistic regression analysis. However, the data typically generated by the transcriptomic technologies creates a need for penalized or modified prospects as a solution to the problems of high dimensionality and overfitting (Turgut, Dagtekin and Ensari, 2018; Morais-Rodrigues *et al.*, 2020; Tabares-Soto *et al.*, 2020; Abapihi *et al.*, 2021). The development of high-end computational algorithms, such as machine learning techniques, has created a new dimension for gene expression analysis.

Machine learning (ML) is an artificial intelligence-based approach that emphasizes building a system that learns automatically from data and improves performance without being explicitly programmed. ML models are trained using a significant amount of data to find hidden patterns required to make decisions (Winston, 1992; Dick, 2019; Micheuz, 2020). Artificial Neural Network (ANN), Classification and regression Trees (CART), Support vector machine (SVM), and vector quantization are some of the architectures used in ML. Recent advancement in the ML domain is deep learning (DL) which is based on artificial neural networks (ANN) (Deng and Yu, 2014; LeCun, Bengio and Hinton, 2015). ANN architectures comprise input, hidden, and output layers of neurons. When more than one hidden layer is used, the ANN method is referred to as the DL method. Basic ML and DL models can work on lower-end machines with less computing power; however, DL models require more powerful hardware to process vast and complex data.

ML techniques, in general, are broadly categorized into supervised and unsupervised learning methods (Jenike and Albert, 1984; Dayan, 1996; Kang and Jameson, 2018; Yuxi, 2018). Supervised learning, which makes use of well-labelled data, is applied for classification and regression analysis. A labelled dataset is used for the training process, which later produces an inferred function to make predictions about unknown instances. Classification techniques train the model to separate the input into different categories or labels (Kotsiantis, 2007). Regression techniques train the model and predict continuous numerical value as an output based on input variables (Fernández-Delgado *et al.*, 2019). Unsupervised techniques, on the other hand, let the model discover information or unknown patterns from the data. We can roughly divide unsupervised learning into clustering and association rules. Clustering used for class discovery is the task of grouping a set of instances in such a way that samples in the same group or cluster are more similar in their properties than the samples in other groups or clusters (Gentleman *et al.*, 2005). Association rules associate links between data instances inside large databases (Kotsiantis and Kanellopoulos, 2006).

The supervised ML techniques have been used for binary classification e.g., identification of cases from control groups in clinical studies, as well as multiclass classification analysis e.g., grading and staging of the disease. ML techniques have been extensively used to analyze gene expression patterns in various complex diseases, such as cancer (Sharma and Rani, 2021), Parkinson's Disease (Peng,

Guan and Shang, 2019), Alzheimer's disease (Kong, Mou and Hu, 2011; Park, Ha and Park, 2020), diabetes (Li, Luo and Wang, 2019), arthritis (Liu *et al.*, 2009; Zhang *et al.*, 2020), etc. The classification algorithms have also contributed to biomarker identification (Jagga and Gupta, 2015), precision treatment (Toro-Domínguez *et al.*, 2019), drug toxicity evaluation (Vo *et al.*, 2020) etc. The unsupervised learning techniques for clustering are routinely used in transcriptomics. The clustering analysis is applied for the study of expression relationships between genes (Liu, Cheng and Tseng, 2011), extracting biologically relevant expression features (Kong *et al.*, 2008), discovering frequent determinant patterns (Prasanna, Seetha and Kumar, 2014), etc.

In supervised and unsupervised learning, the data is subjected to preprocessing, e.g., missing value imputation, normalization, etc. (Figure 2). In supervised learning for classification analysis, the entire dataset is divided into two subsets viz. training and testing/validation. The training dataset, which typically comprises 70-80% of the samples, is used for the construction of a model. The training data can first be subjected to missing value imputation and feature scaling. The preprocessed data is then subjected to feature selection/extraction and model development. The model is then applied to the test/validation dataset, which is also preprocessed in a similar fashion. The preprocessing and feature selection steps are applied to the training dataset after the train-test split to avoid "data leakage". The unsupervised learning which is based on unlabeled data, may include preprocessing steps and data-driven techniques for feature reduction.

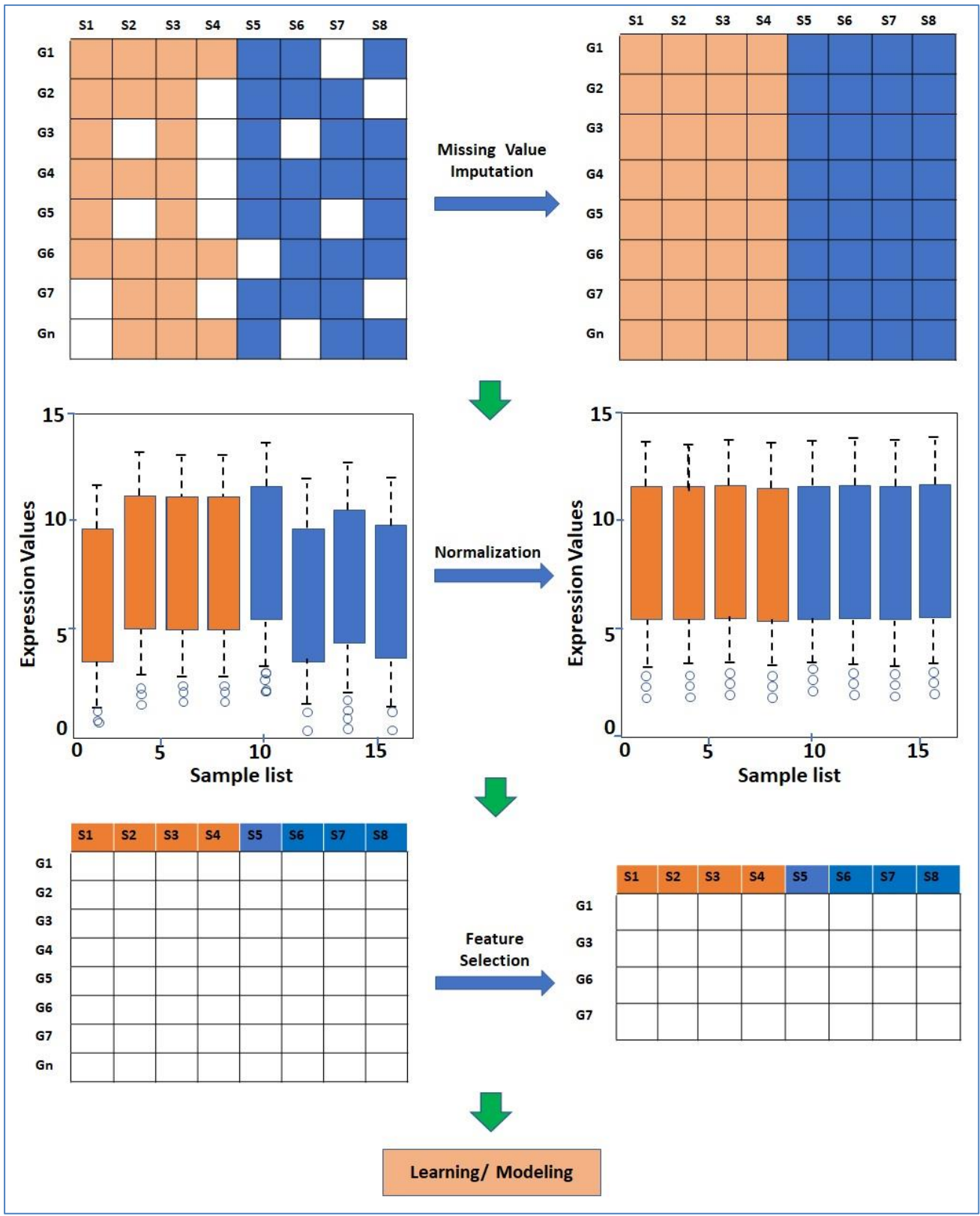

**Figure 2: Steps involved in preprocessing and analysis of gene expression data.** The raw gene expression data may get subjected to missing value imputation, normalization, and feature reduction depending on the type of analysis.

Though missing value imputation, normalization, feature selection, and modelling are important steps in classification analysis, there appears to be very limited literature that reviews them together. Most of the reviews focus either on missing value imputation, features selection, or learning/modelling (Quackenbush, 2001; Dudoit and Fridlyannnd, 2005; Chen *et al.*, 2007; Yip, Amin and Li, 2011; Liew, Law and Yan, 2011; Sahu, Swarnkar and Das, 2011; Khatri, Sirota and Butte, 2012; Tyagi and Mishra, 2013; Bolón-Canedo *et al.*, 2014; Li *et al.*, 2015; Manikandan and Abirami, 2018; Hambali, Oladele and Adewole, 2020; Zhang, Jonassen and Goksøyr, 2021). This creates gaps in understanding of the complete pipeline of the analysis process for researchers from different domains. The objective of this review is to bridge these gaps. Here we discuss various ways to analyze gene expression data and computational methods used at each step. Through this comprehensive review, we also discuss the need for interpretability to provide insights and bring trust to the predictions made. The review is organized into 6 sections. Section 2 broadly covers different missing value imputation approaches along with their advantages and limitations. Section 3 discusses feature scaling techniques applied to gene expression data. In section 4, broad categories of feature selection and dimensionality reduction techniques are discussed. Section 5 covers the different types of gene expression analysis, including class comparison, classification (class prediction), and class discovery. In section 6, we discuss conclusions and future directions.

## 2  Missing value imputation

Gene expression matrices are often riddled with missing gene expression values due to various reasons. In this section, we will discuss sources of missing values and various computational techniques utilized to perform the imputation of missing values. Missing data are typically grouped into three categories: Missing Completely At Random (MCAR), Missing At Random (MAR), and Missing Not At Random (MNAR) (Rubin, 1976; Schafer and Graham, 2002; Aydilek and Arslan, 2013; Mack, Su and Westreich, 2018) (Figure 3). In MCAR, the missing data is independent of their unobserved values and independent of the observed data. In other words, the data is completely missing at random, independent of the nature of the investigation. MAR is a more general class of MCAR where conditional dependencies are accounted for. In MAR, the missingness of data is random but conditionally dependent on observed and unobserved values. In transcriptomics, it can be assumed that all MAR values are also MCAR (Lazar *et al.*, 2016); for example, a channel signal obscured accidentally by a dust particle. In MNAR, the missingness depends on the observed and/or unobserved data. In microarray analysis, values missing due to their low signal intensities are an example of MNAR data.

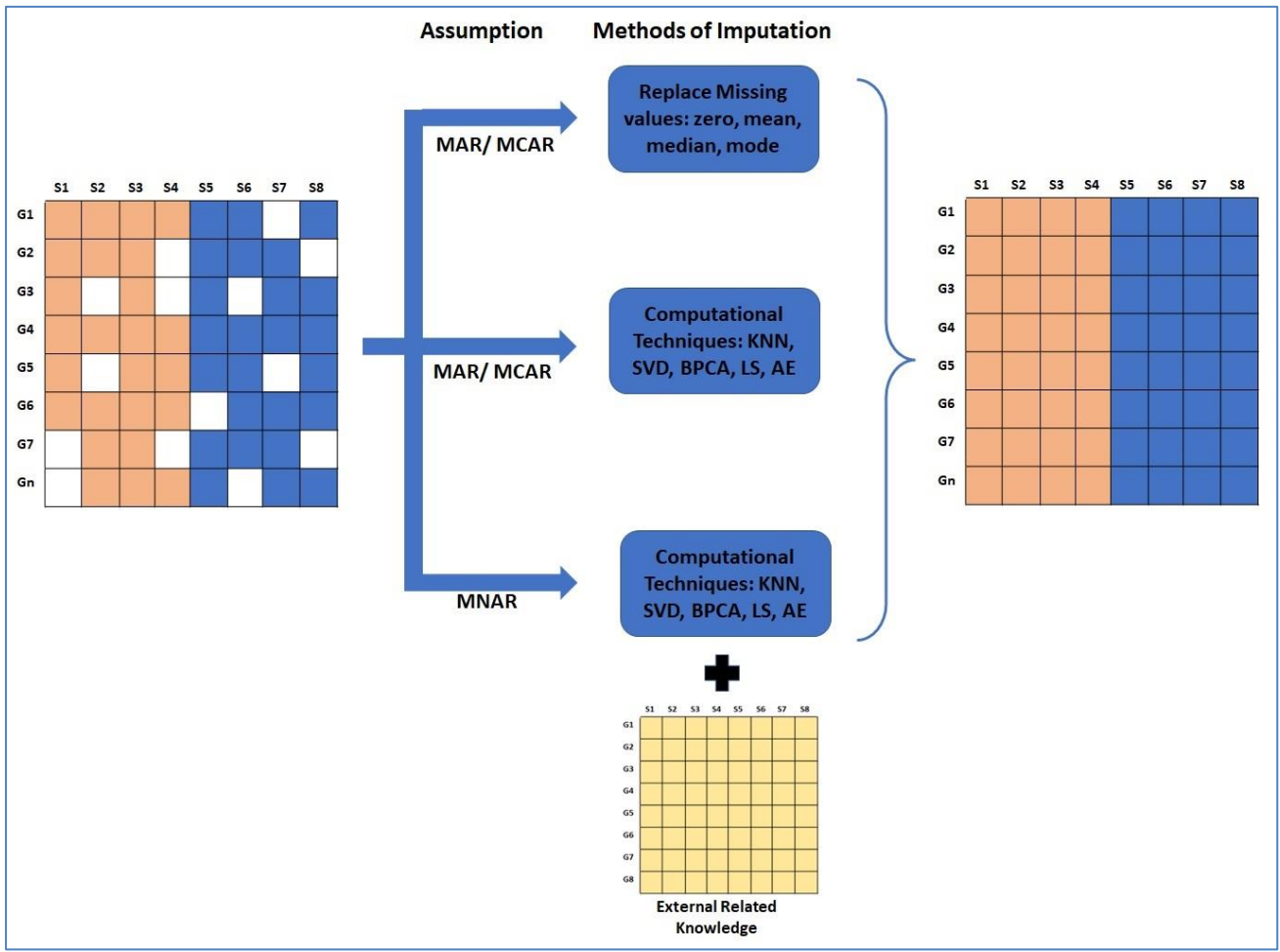

**Figure 3: Types of missing data and methods of imputation for missing values.** The missing values can be Missing Completely At Random (MCAR), Missing At Random (MAR), or Missing Not At Random (MNAR). The missing values are imputed by fixed values; by mean, mode, or median; or by values estimated using advanced computational methods.

MCAR/ MAR and MNAR data can be imputed using two different approaches. In the first method, the missing values are either embedded with a fixed value, or mean, median, or mode. However, this method creates lots of similar values if missing data is high. The second method imputes the missing values using advanced computational techniques. The choice of imputation method depends on the accuracy of the results obtained from the downstream analysis. Computational techniques for estimating missing values can be categorized into four different approaches: Global, Local, Hybrid, and Knowledge Assisted (García-Laencina, Sancho-Gómez and Figueiras-Vidal, 2008; Moorthy *et al.*, 2019; Farswan *et al.*, 2020) (Table 2).

**Global approaches**

Global approaches assume homogeneity of data (e.g. data generated from the same platform and same protocols) and use global correlation information extracted from the entire data matrix to estimate missing values. The Bayesian framework for Principal Component Analysis (BPCA) is based on a probabilistic model that can handle large variations in the expression matrix (Jörnsten *et al.*, 2005; Souto, Jaskowiak and Costa, 2015). In BPCA, the missing value is replaced with a set of random values that are estimated using the Bayesian principle to obtain the relevant principal axes for regression. Singular Value Decomposition (SVD) is another global approach for missing value imputation. SVD is a matrix decomposition method for reducing a matrix to its three constituent parts (Figure 4A). A new matrix that is similar to the original matrix is reconstructed using these constituents in order to reduce noise and impute missing values (Qiu, Zheng and Gevaert, 2018).

Other than above mentioned techniques, ANN-based techniques are also being utilized for the imputation of missing gene expression values. ANN-based methods for imputation include ANNimpute (García-Laencina, Sancho-Gómez and Figueiras-Vidal, 2008), RNNimpute (Bengio and Gingras, 1995), etc. ANNimpute utilizes MLP (Multi-Layered Perceptron) based architecture that is trained with complete observed data (Saha *et al.*, 2017) (Figure 4D). The final weight matrix generated through this process is further used for missing value imputation. RNNimpute utilizes Recurrent Neural Network architecture-based imputation (Bengio and Gingras, 1995) (Figure 4E). Since RNN has feedback connections from its neurons, it can preserve the long-term correlation between parameters.

**Local approaches**

Local approaches utilize a potential local similarity structure to estimate missing values. For heterogeneous data (e.g., data derived from different platforms), the local approach is considered to be very effective. Many local imputation methods have been proposed since 2001. These techniques use a subset of the entire data by estimating underlying heterogeneity. K-Nearest Neighbor (KNN) is a standard ML-based missing-value imputation strategy (McNicholas and Murphy, 2010; Ryan *et al.*, 2010; Pan *et al.*, 2011; Dubey and Rasool, 2021) (Figure 4B). A missing value is imputed by finding the samples closest to the sample from which the gene expression value is missing. It should be noted that a lower number of neighboring points (K) may lead to overfitting of data (Batista and Monard, 2002) whereas a higher K may result in underfitting. Least Square (LS) imputation technique selects a number of most correlated genes using the L2-norm and/or Pearson's correlation (Bo, Dysvik and Jonassen, 2004; Liew, Law and Yan, 2011; Dubey and Rasool, 2021). Support Vector Regression (SVR) method is a non-linear generalization of the linear model used for the imputation of missing gene expression values (Wang *et al.*, 2006; Oladejo, Oladele and Saheed, 2018) (Figure 4C). A significant advantage of the SVR model is that it requires less computational time than other techniques mentioned above (Wang *et al.*, 2006). However, the change in the missing data patterns and the high fraction of missing data limits the effects of SVR. Gaussian Mixture Clustering (GMC) is another technique used for the imputation of missing values that works with highly observable data (Ouyang, Welsh and Georgopoulos, 2004).

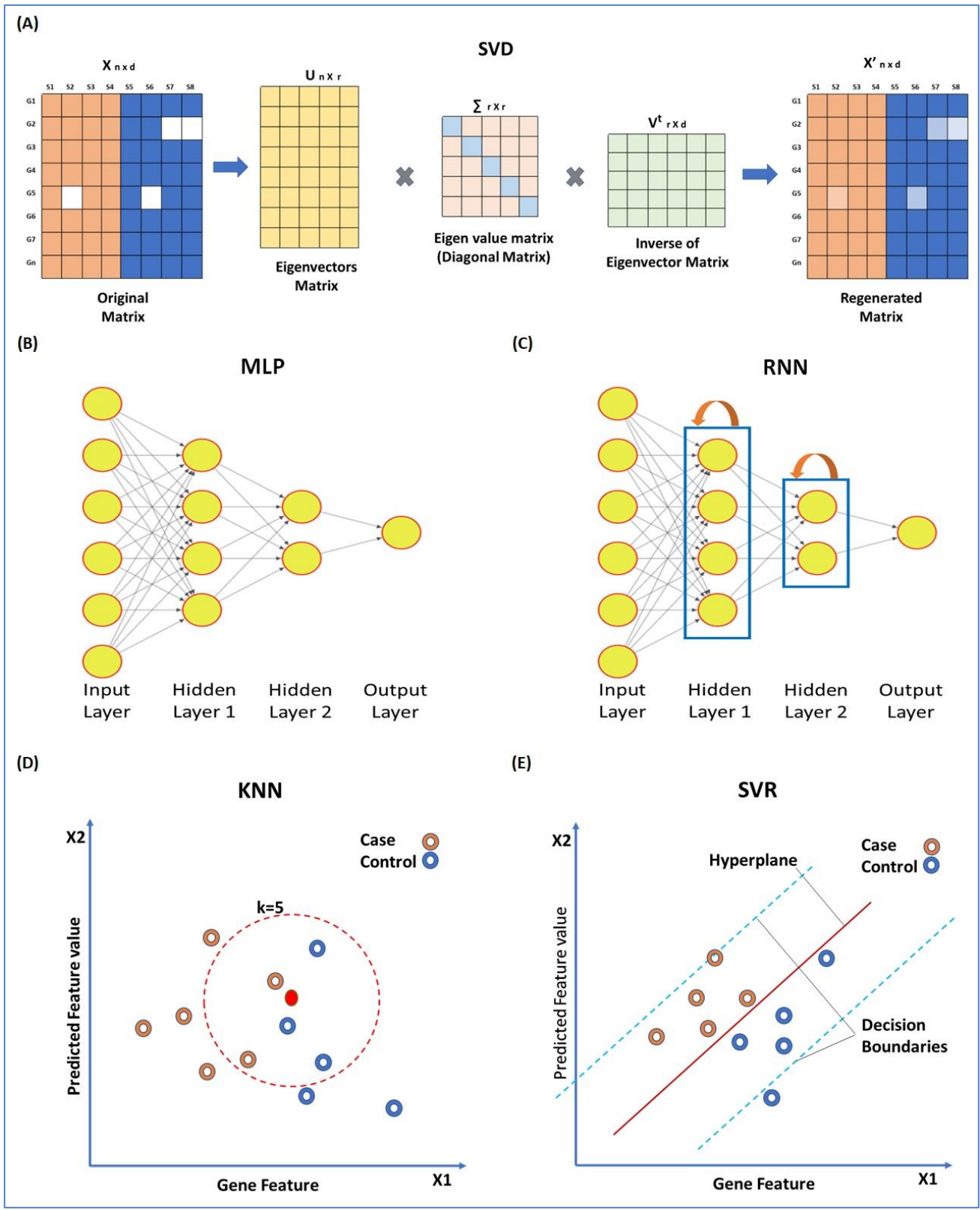

**Figure 4: Approaches for missing value imputation.** Global approaches such as A) Singular value decomposition (SVD), B) Multi-layered perceptron (MLP), and C) Recurrent Neural Network use global correlation information extracted from gene expression array to estimate missing values. Local

approaches such as D) K-nearest neighbor (KNN) and E) Support vector regression (SVR) utilize local similarity in the gene expression data. Hollow circles in D) and E) represent samples.

Some studies have compared the global and local approaches for their performances. SVD and KNN require re-computation of a matrix for every missing value, which results in prolonged evaluation time (Aghdam *et al.*, 2017). SVR, BPCA, and LS try to mine the hidden pattern from the data and seem to perform better than SVD and KNN (Sahu, Swarnkar and Das, 2011) (Tuikkala *et al.*, 2008; Subashini and Krishnaveni, 2011; Qiu, Zheng and Gevaert, 2020).

**Hybrid approach**

The internal correlation among genes affects the homogeneity and heterogeneity of data and, therefore, the performance of global and local imputation approaches (Liew, Law and Yan, 2011). In order to cover both homogeneous and heterogeneous data, a hybrid approach can be very effective. LinCmb is one such hybrid approach for data imputation. LinCmb (Jörnsten *et al.*, 2005) puts more weight on local imputation if data is heterogeneous and has fewer missing values. In contrast, it puts more weight on global methods if data is homogeneous with higher missing values. LinCmb takes an ensemble of row mean, KNN, SVD, BPCA, and GMC. When evaluated, LinCmb's performance was found to be better than each technique it has ensembled. Ensemble missing data imputation method EMDI is another hybrid imputation approach composed of BPCA, matrix completion, and two types of LS and KNN estimators (Pan *et al.*, 2011). It utilizes high-level diversity of data for the imputation of missing values. Recursive Mutual Imputation (RMI) is also a hybrid approach that comprises BPCA and LS to exploit global and local structures in the dataset, respectively (Li *et al.*, 2015). ANN-based Autoencoder (AE) (Qiu, Zheng and Gevaert, 2018) technique consists of encoder, and decoder layers. The encoder converts the input into the hidden representation and the decoder tries to reconstruct the input from the hidden representation. Hence, AE aims to produce output close to the input (García-Laencina, Sancho-Gómez and Figueiras-Vidal, 2008).

**Knowledge-assisted approaches**

Knowledge-assisted approaches incorporate domain knowledge or external information into the imputation process. These approaches are applied when there exists a high missing rate, noisy data, or a small sample size. The solution obtained through this approach is not dependent on the global or local correlation structure that exists in the data but on the domain knowledge. Commonly used domain knowledge includes samples information such as experimental conditions, clinical information, and gene information which includes gene ontology, epigenetic profile, etc. Integrative MISSing Value Estimation (iMISS) (Hu *et al.*, 2006) is one such knowledge-assisted imputation technique. iMISS incorporates knowledge from multiple related microarray datasets for missing value imputation. It obtains coherent neighbors set of genes for every gene with missing data by considering reference data sets. GOImpute (Tuikkala *et al.*, 2006) is another knowledge-assisted imputation technique that uses GO database for knowledge assistance. This method integrates the semantic similarity in the GO with the expression similarity estimated using the KNN imputation algorithm. Projection onto convex sets impute (POCSimpute) (Gan, Liew and Yan, 2006) formulates every piece of prior knowledge into a corresponding convex set to capture gene-wise correlation, array-wise correlation, and known biological constraint. After this, a convergence-guaranteed iterative procedure is used to obtain a solution in the intersection of all these sets. HAIimpute (Xiang *et al.*, 2008) utilizes epigenetic information e.g. histone acetylation knowledge for the imputation of missing values. First, it uses the mean expression values of each gene from each cluster to form an expression pattern. It obtains missing values in the sample by applying linear regression as a primary imputation and uses KNN or LS for secondary imputation. Since knowledge-based methods strongly rely on domain-specific knowledge, they may fail to estimate missing values from under-explored cases with low knowledge available (Wang *et al.*, 2019).

Although a large number of missing value imputation methods are available to the users, there are still quite a few challenges when it comes to the application of imputation methods to the data. Firstly, there is only limited knowledge on the performance of different imputation methods on different types of missing data. The performance of the imputation methods may vary significantly depending on the experimental settings. Therefore, it is important to systematically evaluate the existing methods for their performance on different platforms and experimental settings (Aittokallio, 2009). Secondly, despite the many recent advances, better imputation algorithms that can adapt to both global and local characteristics of the data are still needed. Thirdly, the knowledge-based approaches can also be hybridized with local and/or global approaches to data imputation. More sophisticated algorithms which handle this combinatorial information may work better on the dataset with a higher rate of missing values and can be expected to perform better than those working on transcriptomics data alone (Liew, Law and Yan, 2011).

## 3    Data normalization

Once the missing values are imputed, the datasets can be subjected to downstream analysis. Efficacy of some of the classification methods, e.g., tree-based techniques, linear discriminant analysis, naïve Bayes, etc., does not get affected by variability in the data. However, the performance of class comparison, class discovery, and classification methods, e.g., KNN, SVM etc., may get affected due to technical variations in gene expression signals. The gene expression signals may vary from sample to sample due to technical reasons such as the efficiency of labeling, amount of RNA, and platform used for the generation of data. It is important to reduce the variability due to technical reasons but preserve the variability due to biological reasons. This can be achieved using data normalization or scaling techniques (Brown *et al.*, 1999) (Table 3).

Quantile normalization  (Bolstad *et al.*, 2003; Hansen, Irizarry and Wu, 2012) is a global mean or median technique utilized for the normalization of single channel expression array data. It arranges all the expression values of samples in order, takes average across probes, substitutes probe intensity with average value, and goes back to the original order. Low computational cost is the advantage of quantile normalization. Robust Multi-chip Average (RMA) is a commonly used technique to generate an expression matrix from Affymetrix data (Gautier *et al.*, 2004) or oligonucleotide microarray (Carvalho and Irizarry, 2010). RMA obtains background corrected, quantile normalized gene expression values (Irizarry *et al.*, 2003). Robust Spline Normalization (RSN) used for Illumina data also makes use of quantile normalization (Du, Kibbe and Lin, 2008). Quantile normalization is also used for single color Agilent data(Smyth, 2005). Loess is a local polynomial regression-based approach which can be utilized to adjust intensity levels between two channels (Yang *et al.*, 2002; Smyth and Speed, 2003; Bullard *et al.*, 2010; Baans *et al.*, 2017). Loess normalization performs local regression for each pair of arrays which are composed of the difference and average of the log-transformed intensities derived from the two channels. Two color Agilent data (Smyth, 2005; Du, Kibbe and Lin, 2008) use loess normalization. Log-transformation is the simplest and very common data normalization technique applied to gene expression data (Pochet *et al.*, 2004; Li, Suh and Zhang, 2006; Aziz *et al.*, 2017). This method does not shuffle the relative order of expression values, therefore, does not affect the rank-based test results. Log transformation is often applied to data subjected to prior normalization by other methods such as quantile and loess.

Standardization is a normalization technique that does not bind values to a specific range. Standardization is commonly applied by subtracting the mean value from each expression value. Z-score is one of the most frequently used methods of standardization. The Z-score transformation modifies expression values such that the expression value of each gene is denoted as a unit of standard deviation from the normalized mean of zero (Cheadle *et al.*, 2003). The standardization can also be used with the median instead of the mean (Pan, Lin and Le, 2002). The use of the median is more robust against outliers. Standardization techniques are often used for data visualization.

Feature normalization can have positive and negative effects on the expression array analysis results. It lowers the bias but also decreases the sensitivity of the analysis (Freyhult *et al.*, 2010). Existing normalization methods for microarray gene expression data generally assume a similar global expression pattern among samples being studied. However, scenarios of global shifts in gene expressions are dominant in the datasets of complex diseases, for example, cancers which makes the assumption invalid. Therefore, when applying it should be kept in mind that normalization techniques such as RMA or Loess may arbitrarily flatten the differences between sample groups which may lead to biased gene expression estimates.

## 4      Feature Selection and Feature Extraction

High dimension data often results in the sparsity of information which is less reliable for prediction analysis. As a result, feature selection or feature extraction techniques are typically used to find informative genes and resolve the curse of dimensionality. The dimensionality reduction not only speeds up the training process but also helps in data visualization. Dimensionality reduction is achieved by either selection or extraction of features by transforming the original set of features into new ones. Dimensionality reduction serves as an important step in classification and class discovery analysis. For classification, the dataset is split into training and testing sets, and feature selection/extraction is carried out on the training set. Feature selection and extraction techniques are broadly divided into four categories: filter methods, wrapper methods, embedded methods, and hybrid methods (Tyagi and Mishra, 2013; Dhote, Agrawal and Deen, 2015; Almugren and Alshamlan, 2019) (Figure 5) (Table 4).

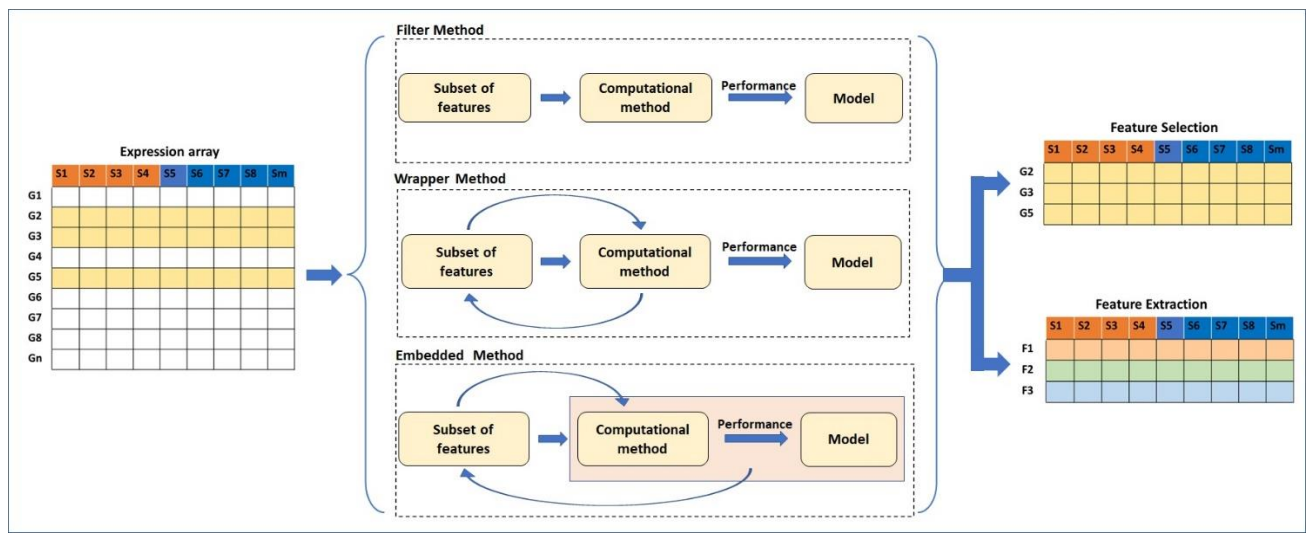

**Figure 5: Methods of feature reduction.** The filter methods of feature selection and extraction work independent of the performance of the learning algorithm. Feature wrapper methods select features

based on the performance of the learning algorithm. Embedded methods perform feature selection as a part of learning process.

**Filter methods**

The filter methods are independent of the performance of the learning algorithm. Statistical methods such as ANOVA, chi-square, t-test, etc. (Pan, Lin and Le, 2002; Saeys, Inza and Larrañaga, 2007; Land *et al.*, 2011; Önskog *et al.*, 2011; Kumar *et al.*, 2015) which are often used for class comparison are also used for the feature selection for prediction analysis. The fold change or p-value is often used as a cutoff parameter for the selection of features. Correlation-based unsupervised learning algorithms are also used for the features selection process (Figure 6A). In correlation-based features selection (CFS), Pearson's coefficient is utilized to compute the correlation among feature genes (Al-Batah *et al.*, 2019). As a next step, the network of genes that has a moderate to high positive correlation with the output variable is retained. Statistical approaches have also been coupled with correlation analysis for feature selection on Maximum Relevance and Minimum Redundancy (MRMR) principles (Radovic *et al.*, 2017). MRMR is a filter approach that helps to achieve both high accuracy and fast speed (Ding and Peng, 2005; Abdi, Hosseini and Rezghi, 2012). The method selects genes that correlate with the condition but are dissimilar to each other. Another commonly used tool is Weighted Gene Co-expression Network Analysis (WGCNA) (Langfelder and Horvath, 2008). This approach is utilized to find the correlation patterns in gene expression across samples as an absolute value of Pearson's correlation (Langfelder and Horvath, 2008). WGCNA groups genes into clusters or modules depending on their co-expression patterns (Agrahari *et al.*, 2018). The eigenvectors generated through clustering can be thought of as a weighted average expression profile, also called eigengenes. These eigengenes can be used to study the relationship between modules and external sample traits. WGCNA is used more often in class comparison analysis for the identification of "hub" genes associated with a trait of interest. Another correlation-based technique, Fast Correlation Feature Selection (FCFS) utilizes a predominant correlation to identify relevant features and redundancy among them without pairwise correlation analysis (Yu and Liu, 2003) (Figure 6B).

The entropy-based methods are supervised learning methods that are used for feature selection. The entropy-based method selects features such that the probability distribution function across external traits have the highest entropy. Information Gain (IG) is a commonly used entropy-based method for feature selection applied to expression array data (Nikumbh, Ghosh and Jayaraman, 2012; Bolón-Canedo *et al.*, 2014; Ayyad, Saleh and Labib, 2019). IG calculates the entropy of gene expression for the entire dataset. The entropy of gene expression for each external trait is then calculated. Based on entropy values, the information gain is calculated for each feature. Ranks are assigned to all the features and a threshold is used to select the features genes. The information gained is provided to the modeling algorithm as heuristic knowledge.

Feature extraction methods are multivariate in nature and are capable of extracting information from multiple feature genes. Classical Principal Component Analysis (PCA), an unsupervised linear transformation technique has been used for dimensionality reduction (Jolliffe, 1986; Pochet *et al.*, 2004; Ringnér, 2008; Adiwijaya *et al.*, 2018) (Figure 6C). PCA builds a new set of variables called principal components (PCs) using original features. To obtain principal components, PCA finds linear projection of gene expression levels with maximal variance over a training set. The PCs with the highest eigenvalues which explain the most variance in data are usually selected for further analysis. Independent component analysis (ICA), another unsupervised transformation method, generates a new set of features from the original ones by assuming them to be linear mixtures of latent variables (Lee and Batzoglou, 2003; Zheng, Huang and Shang, 2006). All features generated using ICA are considered to be statistically independent and hence equally important. As a result, unlike PCA, all components from ICA are used for further analysis. (Hyvärinen, 2013), however, as compared to PCA, ICA is slower. Linear Discriminant Analysis (LDA), on the other hand, is a supervised linear transformation feature reduction method that takes class labels into account and maximizes the separation between classes (Guo and Tibshirani, 2007; Sharma *et al.*, 2014) (Figure 6C). The projection vectors are generated from original features. The projection vectors corresponding to the highest eigenvalue are used for downstream analysis. Similar to PCA, LDA also uses second order statistics. However, as compared to PCA and ICA, LDA offers faster speed and scalability.

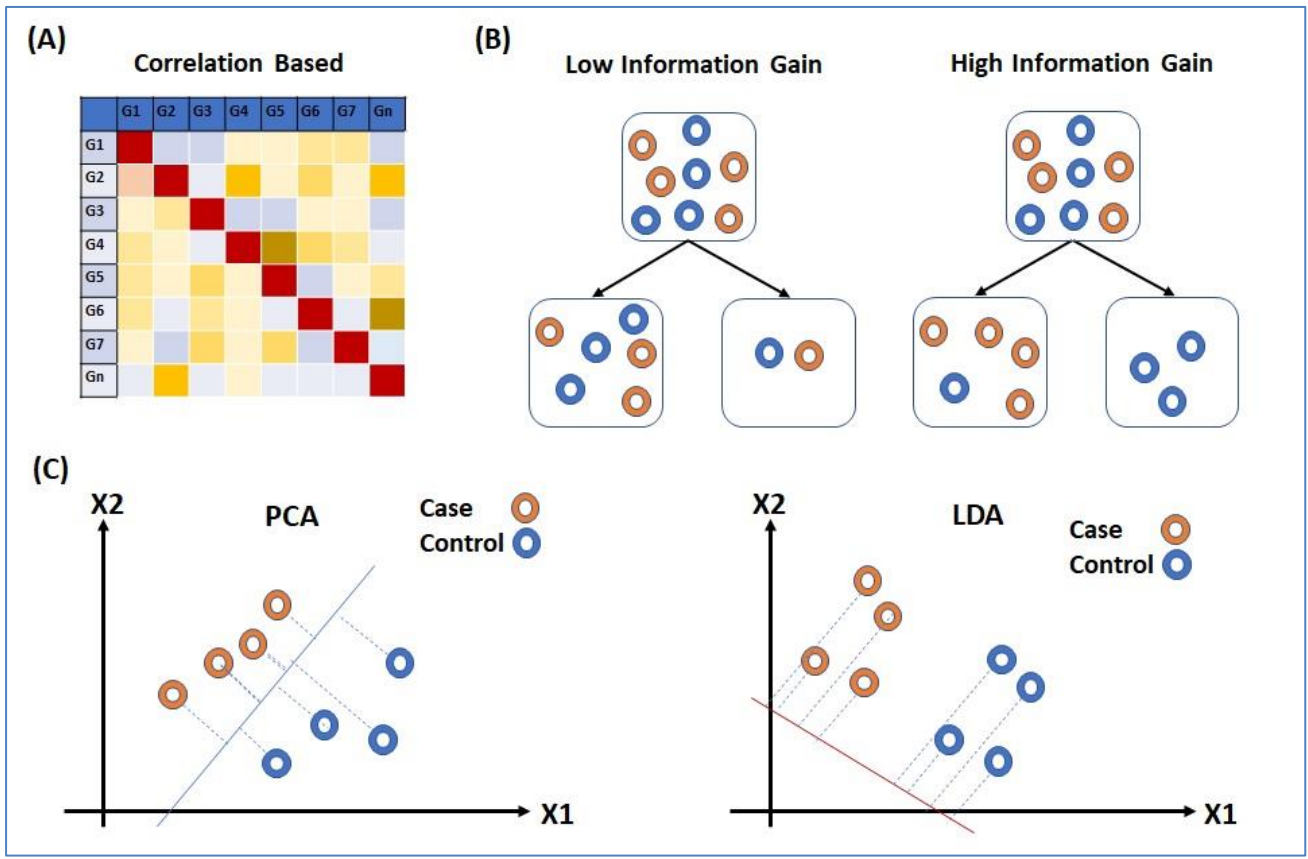

**Figure 6: Filter approaches for feature reduction.** A) Correlation based feature selection (CFS) and B) Information Gain (IG) are feature selection approaches for feature reduction. C) Principal Component Analysis (PCA) and Linear Discriminant Analysis (LDA) extract information from multiple feature genes. Hollow circles in B) and C) represent samples.

All filter approaches (both simple filter and feature extraction methods) ignore the interface with classifier which can result in poor classification performance. This limitation can be overcome by wrapper and embedded approaches.

**Wrapper approach**

The wrapper approach is a feature selection approach that wraps a specific machine learning technique applied to fit the data (Figure 7). The wrapper approach overcomes the limitation of the filter approach by selecting a subset of features and evaluating them based on the performance of the learning algorithm. The process of feature selection repeats itself until the best set of features is found.

Sequential Forward Selection (SFS) is an iterative method of feature selection (Figure 7A). It calculates the performance of each feature and starts with the best performing feature. It then keeps adding a feature with each iteration and keeps checking the performance of the model. A set of features that will produce the highest improvement will be retained, and others will be discarded (Park, Yoo and Cho, 2007; Fan, Poh and Zhou, 2009). Sequential Backward Elimination (SBE), on the other hand, initiates the feature selection process by including all the features in the first iteration and by removing one feature with each iteration (Figure 7B). The effect of elimination of each feature is evaluated based on the prediction performance (Guyon *et al.*, 2002; Dhote, Agrawal and Deen, 2015). Selection or elimination of features in SFS and SBE is based on a scoring function, e.g., p-value, r-square, or residuals sum of squares of the model to maximize performance. A Genetic Algorithm (GA) is a stochastic and heuristic search technique used to optimize a function based on the concept of evolution in biology (Pan, Zhu and Han, 2003) (Figure 7C). Evolution works on mutation and selection processes. In GA, the Information Index Classification (IIC) value for each gene feature is calculated. The IIC value for the feature gene represents its prediction power. As a first step, top gene features with high IIC values are selected for further processing. The selected feature genes are randomly assigned a binary form (0 or 1) to represent a 'chromosome'. A set of chromosomes of the select genes with randomly assigned 0s and 1s creates a 'chromosome population'. The fitness power of each chromosome is calculated by considering only the genes which are assigned a value of 1. 'Fit' chromosomes are selected using techniques such as Roulette-wheel selection, rank selection, tournament selection, etc. The select set of chromosomes is subjected to crossover or mutagenesis to generate the offspring. Upon crossover and mutagenesis, the chromosomes exchange or mutate their information contents. The offspring chromosomes are used for further downstream analysis (Aboudi and Benhlima, 2016; Sayed *et al.*, 2019). There are quite a few variants of GAs to handle the feature selection problem (Liu, 2008, 2009; Ram and Kuila, 2019; Sayed *et al.*, 2019). Other stochastic and heuristic methods are Artificial Bee Colony (ABC) (Li, Li and Yin, 2016), Ant Colony Optimization (ACO) (Alshamlan, Badr and Alohali, 2016), Particle Swarm Optimization (PSO) (Sahu and Mishra, 2012), etc.

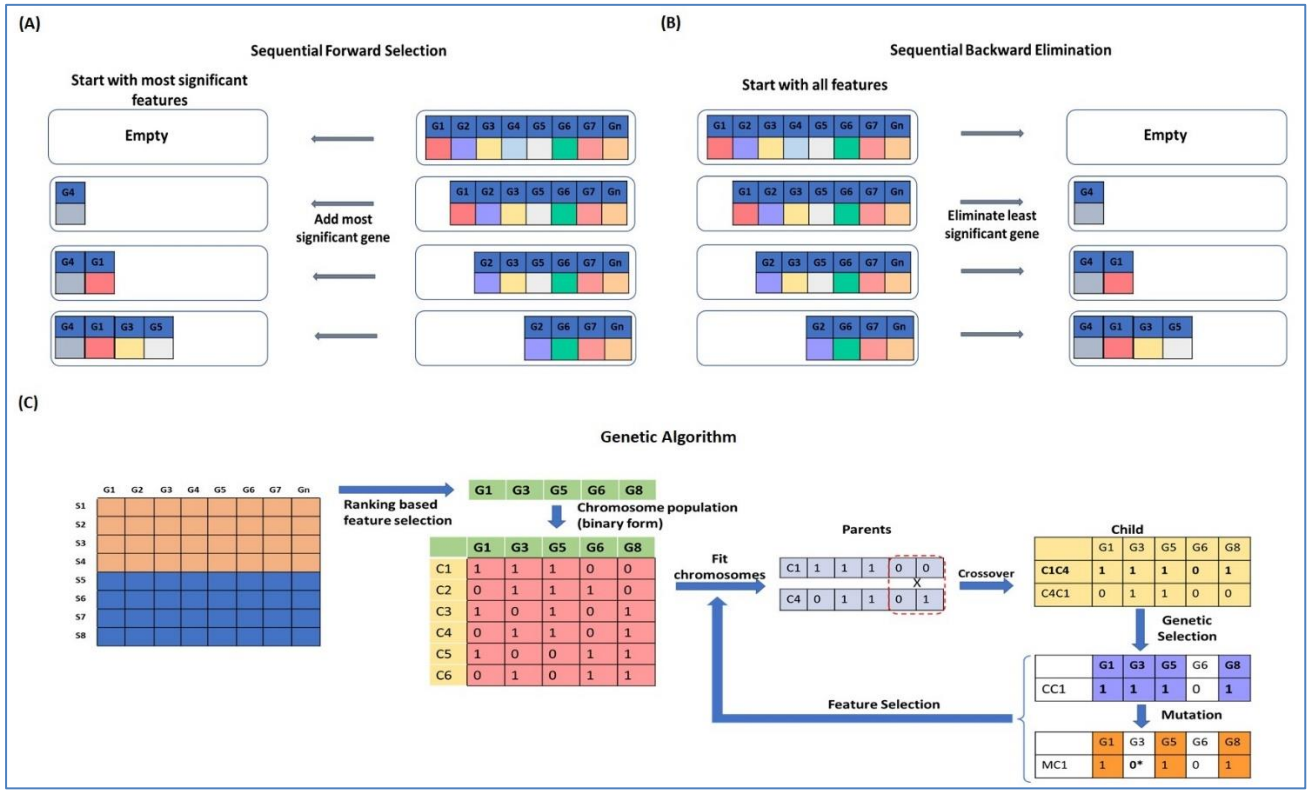

**Figure 7: Wrapper approaches for feature selection.** A) Sequential Forward Selection (SFS), B) Sequential Backward Elimination (SBE) are iterative methods of feature selection. C) Genetic Algorithm (GA) is a stochastic and heuristic search technique that can be used for feature selection.

Though, the wrapped methods provide optimized prediction results as compared to the filter methods they are computationally expensive. This limitation of wrapped methods is addressed by the embedded methods.

**Embedded approaches**

The embedded approaches perform feature selection as a part of the learning process and are typically specific to the learning algorithm. They integrate the importance of both wrapper and filter methods by including feature interaction at a low computational cost. The embedded approach extracts the most contributing features from iterations of training. Commonly used embedded techniques for feature selection are LASSO (Least Absolute Shrinkage and Selection Operator) and Ridge regression (Figure 8A). Both these techniques are regularized versions of multiple linear regression and can be utilized for feature selection (Tibshiranit, 1996). These techniques perform feature selection by eliminating weights of the least important features (Hoffmann, 2007; Ma, Song and Huang, 2007; Meier, Van De Geer and Bühlmann, 2008; Algamal and Lee, 2015). Other than LASSO and Ridge Regression, K-means clustering, SVM, Random Forest and ANN-based techniques are also used.

The K-means clustering technique is an unsupervised method that is utilized to eliminate redundancy in high-dimensional gene expression data (Aydadenta and Adiwijaya, 2018) (Figure 8B). In K-means clustering, an arbitrary K number of points from the data are selected as centroids, and all the genes are allocated to the nearest centroid (MacQueen, 1967; Kanungo *et al.*, 2002). After clustering, a scoring algorithm such as Relief (Kira and Rendell, 1992) is utilized and high-scoring gene features of each cluster are selected for further analysis. The computational complexity of K-means is linear with respect to the number of instances, clusters, and dimensions. Though it is one of the fastest clustering techniques, it may also lead to an incorrect result due to convergence to a local minimum. SVM is a supervised method that is very effective for informative feature selection from data (Guyon, Matin and Vapnik, 1996). SVM generates multiple hyperplanes that separate classes with high prediction accuracy. Compared with several other feature selection techniques such as PCA, LDA, K-means, etc., SVM has shown better prediction performance (Nikumbh, Ghosh and Jayaraman, 2012; Sun, Peng and Shakoor, 2014). The Random Forest (RF) is a supervised approach applied to obtain very small sets of non-redundant genes by preserving predictive accuracy (Díaz-Uriarte and Alvarez de Andrés, 2006; Moorthy and Mohamad, 2012) (Figure 8C). RF is an ensemble of decision trees constructed by randomly selecting data samples from the original data (Breiman, 2001). The final classification is obtained by combining results from the decision trees passed by vote. The bagging strategy of RF can effectively decrease the risk of overfitting when applied to large dimension data. RF can also incorporate connections among predictor features. The prediction performance of RF is highly competitive when compared with SVM and KNN. An important limitation of RF is that many trees can make the model very slow and unproductive for real-time predictions.

ANN-based Autoencoders (AE) (Kramer, 1991) is an unsupervised encoder and decoder technique (Figure 8D). It tries to obtain output layer neuron values as close as possible to input layer neurons using lower-dimensional layers in between. AE can obtain both linear and nonlinear relationships from the input information. AE such as Denoising Autoencoders (DAE)(Vincent and Larochelle, 2008), Stacked Denoising Autoencoder (SDAE) (Vincent *et al.*, 2010; Danaee, Ghaeini and Hendrix, 2017) are utilized to extract functional features from expression arrays and are capable of learning from the dense network. Convolutional Neural Network (CNN) is another ANN-based architecture that is utilized for the feature extraction process in order to improve classification accuracy (Zeebaree, Haron and Abdulazeez, 2018; Almugren and Alshamlan, 2019) (Figure 8E). CNN can extricate local features from the data (LeCun *et al.*, 1998; O'Shea and Nash, 2015). The convolutional layer of CNN extracts the high-level features from the input values. The pooling layer is utilized to reduce the dimensionality of feature maps from the convolution layer.

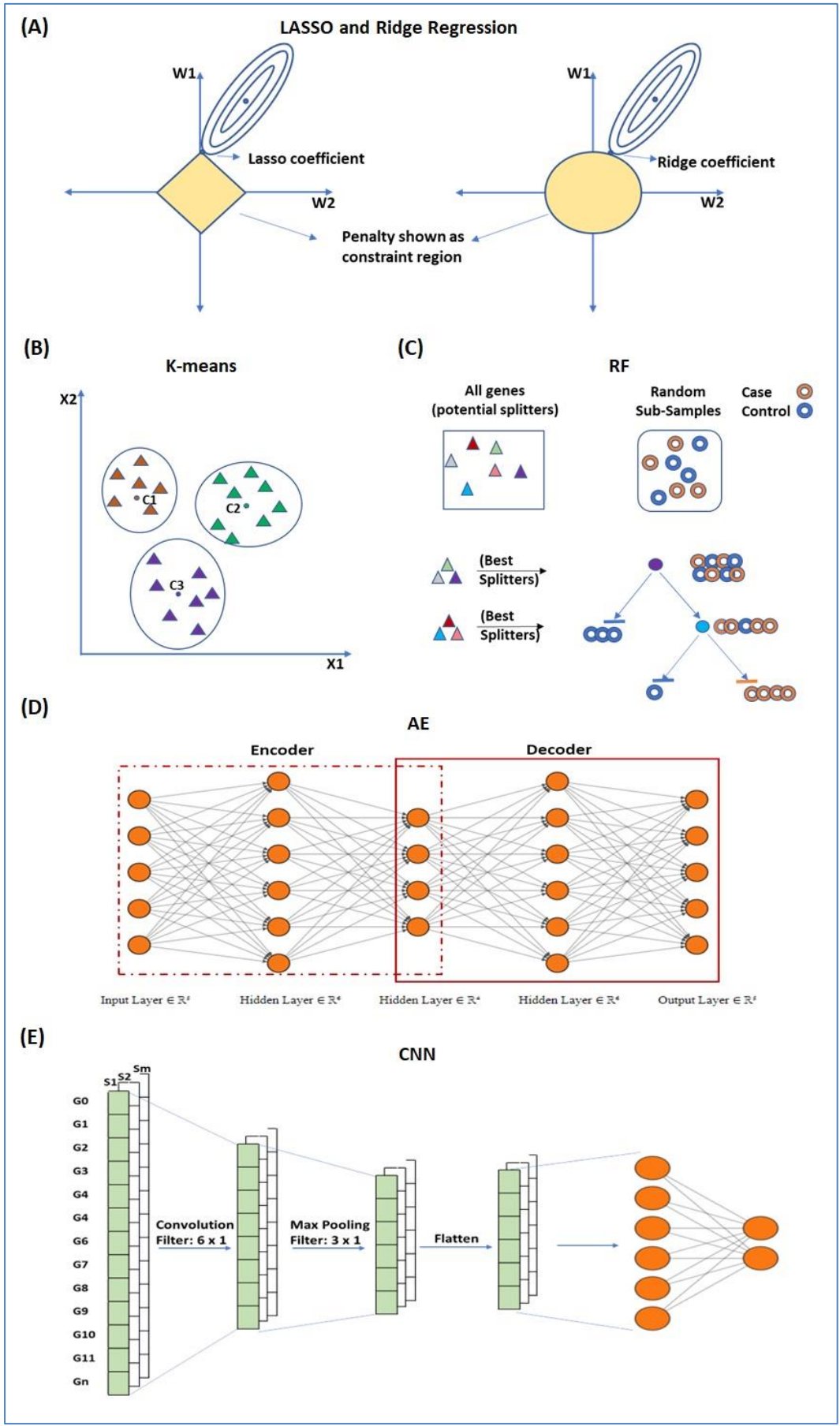

**Figure 8: Embedded approaches performs feature selection and extraction.** A) LASSO and Ridge are regularized versions of multiple linear regression used for feature selection. B) K-means clustering is an unsupervised method for dimensionality reduction that selects feature genes allocated to the nearest centroid. C) Random Forest (RF) is an ensemble of decision trees. D) Convolutional Neural network (CNN) and E) Autoencoders (AE) are deep learning-based methods of feature reduction. Hollow circles in C) represent samples, and solid triangles in B) and C) represent genes.

**Hybrid approach**

A hybrid approach is considered as a combination of two or more filter and wrapper methods. It can reduce the error rate and the risk of overfitting. A well-known feature selection hybrid approach is Recursive Feature Elimination with a linear SVM (SVM-RFE) (Guyon *et al.*, 2002). SVM-RFE utilizes SVMs classification capability and, from the ranked list, recursively deletes the least significant features. This method was taken as a benchmark feature selection method due to its performance. However, its main disadvantage is that it ignores the correlation hidden between the features and requires high computational time (Li, Xie and Liu, 2018). A combination of the mutual information maximization (MIM) and the adaptive genetic algorithm (AGA) has also been proposed for feature selection (Lu *et al.*, 2017). MIM is able to select the advanced feature subset, and AGA speeds up the search in the identification of the substantial feature subsets. This combination of methods is more efficient and robust compared to the individual component (Lu *et al.*, 2017). This technique streamlines the feature selection procedure without getting into classification accuracy on the reduced dataset. MIMAGA-Selection technique can reduce datasets with the number of genes up to 20,000 to below 300 with high classification accuracies. It also removes redundancy from the data. This hybridization approach results in a lower error rate (Bolón-Canedo *et al.*, 2014). This technique is an iterative feature reduction technique. Therefore, with an increase in the size of the microarray dataset, the computational time increases. Co-ABC is a hybrid approach for feature selection based on the correlation Artificial Bee Colony (ABC) algorithm (Alshamlan, 2018). The first step utilizes correlation-based feature selection to filter noisy and redundant genes from high dimensionality domains and the second step utilizes ABC technique to select the most significant genes.

Feature selection or feature extraction process can generate high quality data for classification and predication analysis. It should be noted that for classification analysis, feature selection is carried out only on the training dataset. For clinical applications, it should be noted that model interpretation is important, and feature extraction technique may cause the model interpretation challenging as compared to feature selection techniques.

## 5    Modeling/Learning and Analysis

The final step of analysis of microarray gene expression data is statistical analysis and model learning through computational techniques. Methods used for normalization, gene selection and analyses exhibit a synergistic relationship (Önskog *et al.*, 2011). Class Comparison is one of the most common types of gene expression data analysis for the identification of differentially expressed genes (O'Connell, 2003). To solve the class comparison problems most researchers use standard statistical techniques e.g., t-test, ANOVA, etc. (Storey and Tibshirani, 2003). Scoring enrichment techniques such as z-score or odds ratio are hit-counting methods utilized to describe either the pathway or the functional enrichment of a gene list (Curtis, Orešič and Vidal-Puig, 2005). A higher number of hits shows a higher score and represents greater enrichment.

**Classification (Class Prediction)**

Classification is the process of classifying microarray data into categories or systematic arrangement of microarray data into different classes, e.g., cases and controls. For classification analysis, the entire dataset is divided into two subsets, viz. training and testing. The training dataset, which typically comprises 70-80% of the samples, is used for the construction of a model. To improve the efficiency of classification, it is essential to assess the performance of models. A common way to improve the performance of a model during training is to include an additional validation subset (Refaeilzadeh, Tang and Liu, 2009). The validation dataset comprises 10-15% of the total sample observations used for parameter optimization. The remaining samples are used as a testing dataset. (Refaeilzadeh, Tang and Liu, 2009). However, to assess the generalization ability and prevent model overfitting, instead of setting aside a single validation set, k-fold cross-validation can be an effective solution. Various ML algorithms have been used for classification analysis.

K-Nearest Neighbor (KNN) is one of the techniques that can be utilized for the classification of expression arrays (Kumar *et al.*, 2015; Ayyad, Saleh and Labib, 2019). The classification of a sample is achieved by measuring its distance (e.g., Euclidean distance etc.) from all training samples using the distance metric. The performance of KNN is dependent on the threshold of the feature selection method and is subject to the distance function (Deegalla and Bostr, 2007). An increase in sample size has been shown to increase the computational and time complexity of KNN (Begum, Chakraborty and Sarkar, 2015). Another most used classification technique for expression array data is Nearest Shrunken Centroid (NSC) (Tibshirani *et al.*, 2003; Dallora *et al.*, 2017). It calculates the centroid for each class and tries to shrink each of the class centroids toward the global centroid by threshold. A sample is classified into a class whose centroid is nearest to it based on the distance metric. This method can reduce the effects of noisy genes. However, an arbitrary choice of shrinkage threshold is a limitation of NSC.

A Decision Tree (DT) (Safavian and Landgrebe, 1991) approach can also be utilized for the classification of gene expression data (Peng, Li and Liu, 2006; Krętowski and Grześ, 2007; Chen *et al.*, 2014). A decision tree is also a versatile ML technique that can perform classification as well as regression operations (Safavian and Landgrebe, 1991). DT requires less effort for data preparation during preprocessing. However, a slight variation in the input information can result in a significant variation in the optimal decision tree structure. Also, overfitting is a known limitation of the DT models. Random Forest (RF) (Breiman, 2001) is another algorithm used for the classification and regression analysis of gene expression data. RF is an ensemble of decision trees (Statnikov, Wang and Aliferis, 2008; Aydadenta and Adiwijaya, 2018). While Random Forest has lesser chances of overfitting and provides more accurate results, it is computationally expensive and more difficult to interpret as compared to DT.

Another technique that is utilized for classification analysis using expression arrays is an SVM (Brown *et al.*, 2000; Furey *et al.*, 2000; Ben-Hur *et al.*, 2001; Abdi, Hosseini and Rezghi, 2012; Adiwijaya *et al.*, 2018; Turgut, Dagtekin and Ensari, 2018). For complex non-linear data higher degree polynomials can be added to the cost function of SVM. This will increase the combination of a number of features; however, this results in the reduction of computation speed. To overcome this situation, 'kernel trick' is used, which can handle complex non-linear data without the addition of any polynomial features. Various kernel types can be used with SVM, such as linear, polynomial, radial, etc. In some studies, SVMs performed better than DT and ANN-based techniques (Önskog *et al.*, 2011), whereas, in others the performance of SVM was poor (Tabares-Soto *et al.*, 2020)(Motieghader *et al.*, 2017).

Multilayered CNN, a deep learning algorithm typically applied where the data can be visualized as an image (Neubauer, 1998; Collobert and Weston, 2008), has also been proposed for the analysis of microarray data (Zeebaree, Haron and Abdulazeez, 2018). Each neuron is scanned throughout the input matrix, and for every input, the CNN calculates the locally weighted sum and produces an output value. CNN can deal with insufficient data. CNN involves much less preprocessing and can do far better in terms of results as compared to other supervised techniques.

The performance evaluation for classification analysis using classification techniques can be achieved by error rate or accuracy parameters. Root Mean Squared Error (RMSE) or Root Relative Squared Error (RRSE) are examples of error-rate-based evaluation. The accuracy metric is the most common performance evaluation parameter utilized to find the accuracy of classification. However, accuracy alone is not enough for performance evaluation (McNee, Riedl and Konstan, 2006; Sturm, 2013) and therefore, a confusion matrix is computed. A set of predictions is compared with actual targets to compute the confusion matrix. The confusion matrix represents true positives (TP), true negatives (TN), false positives (FP), and false negatives (FN). TP, TN, FP and FN are utilized to calculate more concise metrics such as precision, recall (sensitivity), specificity, Matthew's correlation coefficient (MCC), etc. ROC (Receiver Operating Characteristic) curve and Precision-Recall curve are other standard tools used by binary classifiers as performance measures. ROC and MCC are more robust measures as compared to accuracy since accuracy is affected by class imbalance (Chicco and Jurman, 2020).

The problem of classification of expression data is both biologically important and computationally challenging. From a computational perspective one of the major challenges in analyzing microarray gene expression data is a small sample size. Error estimation is greatly affected by the small sample size, and the possibility of overfitting of data is very high (Hambali, Oladele and Adewole, 2020). Another important issue in gene expression array data analysis is class imbalance for the classification tasks. In clinical research on rare diseases, generally, the number of case samples is very less as compared to healthy controls which may lead to biased results. With decreasing costs of microarray profiling and high-throughput sequencing, this challenge can be expected to be resolved in the near future.

**Class Discovery**

The third type of microarray analysis is class discovery which involves the analysis of a set of gene expression profiles for the discovery of novel gene regulatory networks or sample types. Hierarchical Clustering Analysis (HCA) is a simple process of sorting instances into groups of similar features and is very commonly used for the analysis of expression array data (Eisen *et al.*, 1998). Hierarchical clustering produces a dendrogram which is a binary tree structure and represents the distance relationships between clusters. HCA is a highly structured approach and the most widely used technique for expression analysis (Bouguettaya *et al.*, 2015). However, the graphical representation of hierarchy is very complex in HCA. The lack of robustness and inversion problems complicates the interpretation of the hierarchy. HCA is also sensitive to small data variations. Self-Organizing Maps (SOM) is another clustering technique used for the identification of prevalent gene expression patterns and simple visualization of specific genes or pathways (Tamayo *et al.*, 1999). SOM can perform non-linear mapping of data with a two-dimensional map grid. Unlike HCA, SOM is less sensitive to small data variations (Nikkila *et al.*, 2002).

K-means is an iterative technique that minimizes the overall within-cluster dispersion. K-means algorithm has been utilized to discover transcriptional regulatory sub-networks of yeast without any prior assumptions of their structure (Tavazoie *et al.*, 1999). The advantage of K-means over other clustering techniques is that it can deal with entirely unstructured input data (Gentleman and Carey, 2008). However, the K-means technique easily gets caught with the local optimum if the initial center points are selected randomly. Therefore various modified versions of K-means are applied for converging to the global optimum (Lu *et al.*, 2004; Nidheesh, Abdul Nazeer and Ameer, 2017; Jothi, Mohanty and Ojha, 2019).

Another technique for class discovery analysis is the Bayesian probabilistic framework which uses Bayes theorem (Friedman *et al.*, 2000; Baldi and Long, 2001). This technique is a good fit for small sample sizes of microarray studies; however, it is computationally exhaustive for a dataset with a very high number of samples and features. Nonnegative Matrix Factorization (NMF) is also a clustering technique utilized for pattern analysis of gene expression data (Kim and Tidor, 2003; Brunet *et al.*, 2004). NMF involves factorization into matrices with nonnegative entries and recognizes the similarity between sub-portions of the data corresponding to localized features in expression space (Kim and Park, 2007; Devarajan and Ebrahimi, 2008).

Evaluation measures for clustering algorithms utilized for class discovery can be of three different types, viz. internal validation index, relative validation index, and external validation index (Dalton, Ballarin and Brun, 2009). The internal validation index method calculates properties of the resulting clusters based on internal properties of clusters such as compactness, separation, and roundness. Dunn's Index and Silhouette Index are examples of internal validation indices. The relative validation indexing method compares clusters generated by algorithms with different parameters or subsets of the data. It can measure the stability of the technique against variations in the data, or consistency of the results in the case of redundancy. The figure of merit index and instability index are examples of relative validation indices. External validation index method compares the groups generated by the clustering technique to the actual cluster of the data. Generally, external methods are considered to be better correlated to the actual error as compared to internal and relative indexing methods. Hubert's Correlation, Rand Statistics, Jaccard Coefficient, and Folke's and Mallow's index are a few examples of external evaluation parameters. Table 5 describes all the evaluation parameters discussed above.

While dealing with a very large number of gene features in expression arrays, multiple gene feature selection techniques are available to deal with dimensionality problem. However, an elaborate study is required to identify optimum methods for downstream analysis that can be combined with specific dimensionality reduction techniques.

## 6    Conclusions and future directions

In this paper, we have attempted to describe the complete pipeline for the analysis of expression arrays. Conventional ML methods for missing value imputation, dimensionality reduction, and classification analysis have achieved success. However, with an increase in data complexity, deep learning techniques may find increasing usage. The current applications of clinical research benefit from the data coming from different modalities. For gene expression data analysis of complex diseases, data sparsity or class imbalance is a real concern. This issue can be addressed with the recent technology of data augmentation, for example, Generative Adversarial Networks (GANs) (Chaudhari, Agrawal and Kotecha, 2020). The aim of any clinical research is not only to predict but also to know the reasons behind the decisions made by downstream analysis. This understanding of the undercover mechanism with some evidence makes the model interpretable. Since the microarray data is used in clinics, it is important to develop interpretable models which help to understand the problem and the situation where the model may fail (Holzinger *et al.*, 2017). Interpretation models such as perturbation-based, derivative-based, local and global surrogate-based should get attention to solve these problems (Ribeiro, Singh and Guestrin, 2016; Zou *et al.*, 2019).

## 7 Conflict of Interest

The authors declare that the research was conducted in the absence of any commercial or financial relationships that could be construed as a potential conflict of interest.

## 8 Authors' Contributions

## 9 Funding

This work has been supported by the Scheme for Promotion of Academic and Research Collaboration (SPARC) 2018-19, MHRD (project no. P104).

## 10 Acknowledgements

Nikita Bhandari was supported by the Junior Research Fellowship Award 2018 by Symbiosis International Deemed University, India. Satyajeet Khare is also a beneficiary of a DST SERB SRG grant (SRG/2020/001414).